
\magnification=1200
\vsize=7.5in
\hsize=5in
\tolerance 10000

\pageno=0
\def\ep{\delta}
\def\de{\epsilon}
\def\P{{\cal P}}

\def\L{p_0}
\line{\hfil TAUP 1850-90}
\vskip .1in
\line{\hfil April 1993}
\baselineskip 12pt plus 1pt minus 1pt
\centerline{\bf }
\smallskip
\centerline{{\bf MEASUREMENTS, ERRORS, AND NEGATIVE KINETIC ENERGY}}
\vskip 24pt
\centerline{Yakir Aharonov}
\vskip 12pt
\centerline{\it School of Physics and Astronomy, Tel-Aviv University}
\centerline{\it Ramat-Aviv, Tel-Aviv 69978 Israel}
\vskip 8pt
\centerline{and}
\vskip 8pt
\centerline{\it Department of Physics, University of South Carolina}
\centerline{\it Columbia, SC  29208 U.S.A.}
\vskip 12pt
\centerline{Sandu Popescu}
\vskip 12pt
\centerline{\it Universit\'e Libre de Bruxelles}
\centerline{\it Campus Plaine, C.P. 225, Boulevard du Triomphe}
\centerline{\it B-1050 Bruxelles, Belgium}
\vskip 12pt
\centerline{Daniel Rohrlich}
\vskip 12pt
\centerline{and}
\vskip 12pt
\centerline{Lev Vaidman}
\vskip 12pt
\centerline{\it School of Physics and Astronomy, Tel-Aviv University}
\centerline{\it Ramat-Aviv, Tel-Aviv 69978 Israel}
\vskip 1in
\centerline{{\bf Abstract}}
An analysis of errors in measurement yields new insight into the penetration of
quantum particles into classically forbidden regions. In addition to
``physical" values, realistic measurements yield ``unphysical" values which, we
show, can form a consistent pattern.  An experiment to isolate a particle in a
classically forbidden region obtains negative values for its kinetic energy.
These values realize the concept of a {\it weak value}, discussed in previous
works.
\eject
\baselineskip 24pt plus 2pt minus 2pt
\goodbreak
\bigskip
\noindent{\bf 1. \quad Introduction}
\medskip
\nobreak
When the word ``quantum" first entered the language of physics, it meant a
restriction on possible values of energy.  Although the quantum theory that
later emerged has many other aspects, it is still axiomatic that the only
observable values of a physical quantity are the eigenvalues of a corresponding
quantized operator.  The more precise our measurements, the more clearly this
restriction stands out; thus when we obtain values that are not eigenvalues,
we interpret them as errors. Still, measurements are uncertain in practice, and
can even yield classically forbidden, ``unphysical" values. We have uncovered
remarkable regularities in the way that ``unphysical" values can appear in
sequences of measurements, suggesting that these values may not be unphysical
at all.  In quantum theory, it seems, not only are physical quantities not
restricted: they can take values outside the classically allowed range.  Here
we discuss this new effect, and show how it arises in the context of barrier
penetration by quantum particles.

     The phenomenon of barrier penetration, such as tunnelling through a
potential barrier, is an outstanding example of quantum behaviour. Quantum
particles can be found in regions where a classical particle could never go,
since it would have negative kinetic energy.  But in quantum theory, too, the
eigenvalues of kinetic energy cannot be negative.  How, then, can a quantum
particle ``tunnel"? The apparent paradox is resolved by noting that the wave
function of a tunnelling particle only partly overlaps the forbidden region,
while a particle found within the forbidden region may have taken enough energy
from the measuring probe to offset any energy deficit.  There is no wave
function that represents a particle restricted to a region where its potential
energy is larger than its total energy.

    Nevertheless, we will show that actual measurements of kinetic energy can
yield negative values, and that, under proper conditions, a remarkable
consistency appears in these apparent errors. In a model experiment, we measure
the kinetic energy of a bound particle to any desired precision. We then
attempt to localize the particle within the classically forbidden region. The
attempt rarely succeeds, but whenever it does, we find that the kinetic energy
measurements gave an ``unphysical" negative result; moreover, these results
cluster around the appropriate value, the difference between the total and the
potential energy. This consistency, which seems to come from nowhere -- a
background of errors -- suggests strongly that the notion of a quantum
observable is richer than the one generally accepted.  Previous papers
suggesting this conclusion analyze a measurement of spin$^1$
and a quantum time machine.$^2$
\goodbreak
\bigskip
\noindent{\bf 2. \quad Analysis of errors in measurement}
\medskip
\nobreak
We begin by reviewing the standard von Neumann$^3$ theory of measurement in
non-relativistic quantum mechanics. Suppose we wish to measure a dynamical
quantity $C$.  We choose a measuring device with an interaction Hamiltonian
$$H_{int}= g(t) P C~~~~,
\eqno(1)$$
where $P$ is a canonical momentum of the measuring device; the conjugate
position $Q$ corresponds to the position of a pointer on the device. The
time-dependent coupling constant $g(t)$ is nonzero only for a short time
interval corresponding to the measurement, and is normalized so that
$$\int g(t)dt=1~~~~.
\eqno(2)$$
When the time interval is very short, we call the measurement impulsive. For an
impulsive measurement, $H_{int}$ dominates the Hamiltonians of the measured
system and the measuring device.  Then, since $\dot Q = {i \over \hbar}
[H_{int} ,Q] = g(t)C$, we obtain (in the Heisenberg representation) the result
$$Q _{fin}-Q _{in}=C~~~~,
\eqno(3)$$
where $Q _{fin}$ and $Q _{in}$ denote the final and initial settings of the
pointer.

     In an ideal measurement the initial position of the pointer is precisely
defined, say $Q _{in}=0$, and so from its final position we read the precise
value of $C$.  But in practice, measurements involve uncertainty.  To model
a source of uncertainty, we can take the initial state of the pointer to be
$$\Phi _{in} (Q) = (\de^2 \pi )^{-1/4} e^{ -{{Q ^2} /{2\de ^2}}}
\eqno(4)$$
The uncertainty in the initial position of the pointer produces errors of order
$\de $ in the determination of $C$; when $\de \rightarrow 0$ we recover the
ideal measurement. Suppose that the system under study is initially in an
eigenstate of $C$ with eigenvalue $c_i$.  Ideal measurements can yield only the
result $c_i$. But when the pointer itself introduces uncertainty, other results
are possible, indeed a scatter of results, with a spread of about $\de$, and
peaked at the eigenvalue $c_i$. If the measuring device works as described,
then
{\it any} measured value is possible, although large errors are exponentially
suppressed.  There is no mystery in the appearance of such errors; they are
expected, given the uncertainty associated with the measuring device.
Measurements of a positive definite operator such as $p^2$ could even yield
negative values. Of course, the dial of the measuring device might have a pin
preventing negative readings, but let us assume that it does not.  Even if the
negative values themselves are unphysical, they are part of a distribution
representing the measurement of a physical quantity.  They should not be thrown
out, since they give information about the distribution and contribute to the
best estimate of the peak value.

     The standard theory of measurement not only allows errors, it also
prescribes their interpretation:  they constitute scatter around a true
physical value which can only be one of the eigenvalues of the operator
measured.  Of course, the system under study may not be in an eigenstate of the
measured operator.  Then results of measurements will be distributed according
to quantum probabilities, folded with errors due to the measuring device. Since
these errors originate in the measuring device, and not in the system under
study, it seems that they cannot depend on any property of the system. However,
closer analysis of these errors in the context of {\it sequences} of
measurements reveals a pattern which, far from being random, clearly reflects
properties of the system under study. The pattern emerges only after selection
of a particular final state of the system. In the next section, we take a
particular example and analyze it in detail to show how and where the surprise
appears.
\goodbreak
\bigskip
\noindent{\bf 3. \quad Negative kinetic energy}
\medskip
\nobreak
     Our example may be summarized as follows: we prepare a sufficiently large
ensemble of particles bound in a potential well, in an eigenstate of energy,
and measure the kinetic energy of each particle to a given precision. The
results of these measurements are predictably scattered, and even include some
negative values, although the kinetic energy spectrum is positive. Then we
measure the position of each particle and select only those cases where the
particle is found within some region ``far enough" from the well -- with ``far
enough" depending on how precisely the kinetic energy was measured.  In almost
all such cases, we find that the measured kinetic energy was {\it negative}.
Not only are the measured values negative, they also cluster around a
particular negative value appropriate to particles in the classically forbidden
region. Also, the spread of the clustering is the characteristic spread for
kinetic energy measurements with this device.

     We begin with a particle trapped in a potential well.  The Hamiltonian
is$^4$
$$H={{p^2}\over {2m}}+V(x)~~~~,
\eqno(5)$$
with $V(x)=-V_0$ for  $\vert x\vert <a$ and $V(x)=0$ for $\vert x\vert >a$.
We prepare an ensemble of particles in the ground state of this Hamiltonian,
with energy $E_0 <0$:
$$\vert \Psi_{in} \rangle =\vert E_{0}\rangle ~~~~.
\eqno(6)$$
Following von Neumann, we model a measurement of kinetic energy with an
interaction Hamiltonian
$$H_{int}=g(t) P~{p^{2}\over 2m}~~~~,
\eqno(7)$$
where $P$ is a canonical momentum conjugate to the position $Q$ of a pointer on
the measuring device. As in Eq. (2),
we assume that the coupling between the particle and device is turned on so
briefly that the Hamiltonian reduces to $H_{int}$, and we obtain for the
operator $Q$
$$Q _{fin}-Q _{in}={p^{2}\over 2m}~~~~.
\eqno(8)$$
The initial state of the pointer is given by Eq. (4).
The uncertainty in the initial state of the pointer leads to errors of order
$\de $ in the measurement of kinetic energy.

     Initially, the particle and device are in a product state $\Psi_{in} (x)
\Phi _{in} (Q) $; after the interaction is complete, the state is
$$e^{-{i\over \hbar }P~ {{p^{2}} /2m}}
\Psi_{in} (x) \Phi _{in} (Q) ~~~~,
\eqno(9)$$
in which the particle and the device are correlated. Now we consider kinetic
energy measurements followed by a final measurement of position, with the
particle found far outside the potential well. For the final state we choose a
gaussian wave packet with its center far from the potential well,
$$\Psi _{fin} (x) = (\ep^2 \pi )^{-1/4} e^{-(x-x_{0})^{2}/ 2\ep^2 } ~~~~,
\eqno(10)$$
and we require $\delta > \alpha \hbar^2 /m\de$. We can now be more definite
about what it means for the particle to be ``far enough" from the potential
well; the condition on $x_0$ is
$$\alpha {x_0} >>\left( \alpha^2 \hbar^2 /2m \epsilon \right)^2
{}~~~~.\eqno(11)$$
Since $\alpha^2 \hbar^2 /2m = \vert E_0 \vert$, the expression in parentheses
is the ratio of the magnitude of the effect, $\vert E_0 \vert$, to the
precision of the measurement, $\epsilon$. This condition is derived in the
Appendix.  Note that for more precise measurements of kinetic energy $(\de
\rightarrow 0)$, the final state is selected at increasing distances from the
potential well $(x_0 \rightarrow \infty )$.

     The state of the measuring device after the measurement, and after the
particle is found in the state $\Psi_{fin} (x)$, is obtained by projecting the
correlated state of the particle and measuring device onto the final state of
the particle $\Psi _{fin} (x)$. Apart from normalization, the final state of
the measuring device is
$$\Phi _{fin} (Q) = \langle \Psi_{fin} \vert
e^{-{i\over \hbar }P~{{p^{2}} /2m}}
\vert \Psi_{in} \rangle \Phi_{in} (Q) ~~~~,
\eqno(12)$$
where $\Psi_{in} (x) =\vert E_0\rangle $. For simplicity, we take $V(x)$ in Eq.
(5)
to be a delta-function potential ($a \rightarrow 0$). Then $\Psi_{in} (x)$ is
$\sqrt{\alpha} \exp (-\alpha \vert x\vert)$.

     The exponent in Eq. (12)
contains the operators $P$ and $p$.  It is convenient to express $\Psi_{in}
(x)$ via its Fourier transform,
$$
\Psi_{in} (x) = {{\hbar \alpha^{3/2}} \over \pi} \int dp
{{e^{-ipx/\hbar} }\over {\alpha^2 \hbar^2 + p^2 }}
{}~~~~,\eqno(13)$$
and replace the operator $p$ with its eigenvalue.  The exponential of $-iPp^2
/2m\hbar$ effects a translation of $Q$ in $\Phi_{in} (Q)$, and we obtain (up to
a normalizing factor)
$$
\Phi_{fin} (Q) = {\pi \over {\hbar \alpha}}
e^{\alpha x_0 - \alpha^2 \delta^2 /2 }
\int dp {{e^{-p^2  \delta^2 /2\hbar^2 - i p x_0/\hbar} }
\over {\alpha^2 \hbar^2 +p^2} }\Phi_{in} (Q - p^2 /2m)
{}~~~~.\eqno(14)$$
This integral has poles at $p= \pm i\alpha \hbar$; we may evaluate it by
integration on a contour including a line of $p$ with imaginary part $-i\L$,
for any $\L > \hbar \alpha$.  The integral in \hbox{Eq. (14)}
then reduces to two terms:  a pole term
$$
\Phi_{in} (Q+ {{\alpha^2 \hbar^2}   / {2m}})
{}~~~~, \eqno(15)$$
and the integral Eq. (14)
with $p$ replaced by $p-i\L$.  The pole term represents the measuring device
with its pointer shifted to the negative value $-\alpha^2 \hbar^2 /2m$. If the
final state included only this term, measurements would yield $-\alpha^2
\hbar^2 /2m$ for the kinetic energy, up to a scatter $\de$ characteristic of
the measuring device.

     The correction to the pole term is the integral in $p-i\L$,
$$
{{\hbar \alpha } \over \pi} e^{\alpha x_0 -\alpha^2 \delta^2 /2 }
\int_{-\infty}^{\infty} dp
{{e^{-(p-i\L )^2 \ep^2 /2 \hbar^2  -
i(p -i\L )x_0 /\hbar} }\over {\alpha^2 \hbar^2 + (p-i\L )^2} }
\Phi_{in} \left( Q - (p-i\L )^2 / 2m \right)
{}~~~~.\eqno(16)$$
We can bound the magnitude of the correction by replacing the integrand with
its absolute value. The integral over the absolute value converges (see
Appendix). Since we replaced the integrand with its absolute value, the only
dependence on $x_0$ that remains is the exponential $e^{(\alpha -  \L / \hbar
)x_0}$.  Since $\alpha <\L / \hbar $, the correction to $\Phi_{in} (Q+\alpha^2
\hbar^2 /2m)$ can be made arbitrarily small by taking $x_0$ large, as in Eq.
(11).
In this limit, the final state of the measuring device shows the ``unphysical"
result $-\alpha^2 \hbar^2 /2m$ for the kinetic energy.

     We thus obtain a correlation between position measurements and prior
kinetic energy measurements:  nearly all particles found far outside the
potential well yielded negative values of kinetic energy.  On the other hand,
we could look at the entire set of data differently.  We could consider all
particles that produced negative values of kinetic energy, and ask about their
final position. We would find nearly all these particles {\it inside} the well.
The correlation works one way only.  Prior kinetic energy measurements on
particles found far from the well cluster around a negative value, but position
measurements on particles yielding negative values of kinetic energy cluster
around zero.  How do we interpret this one-way correlation?
\goodbreak
\bigskip
\noindent{\bf 4. \quad Interpretation}
\medskip
\nobreak
     Our example suggests that particles in a classically forbidden region have
negative kinetic energy. But the conventional interpretation of quantum
mechanics has no place for negative kinetic energy. Measurements correspond to
eigenvalues or to expectation values only.  These must be positive in the case
of kinetic energy, so negative measured values of kinetic energy must be
errors.

     However, the conventional interpretation involves an assumption about how
measurements are made.  The conventional interpretation considers measurements
on ensembles of systems prepared in an initial state, without any conditions on
the final state of the systems.  Such an ensemble, defined by initial
conditions only, may be termed a {\it pre-selected} ensemble.  By contrast, we
consider measurements made on {\it pre- and post-selected} ensembles, defined
by both initial and final conditions.  The experiment of the previous section
is an example of a measurement on a pre- and post-selected ensemble.  It is
natural to introduce pre- and post-selected ensembles in quantum theory:  in
the quantum world, unlike the classical world,  complete specification of the
initial state does not determine the final state.

     A measurement on a pre- and post-selected ensemble involves a
pre-selection, a measurement, and a post-selection. Aharonov, Bergmann, and
Lebowitz$^5$ (ABL) gave a formula for the result of the intermediate
measurement.  Let an operator $C$ be measured at time $t$ between a
pre-selected state $|a\rangle$ at time $t_1$ and a post-selected state
$|b\rangle$ at time $t_2$.  If $C$ has eigenvalues $c_j$, then the probability
$\P (c_j )$ that the intermediate measurement of $C$ yields $c_j$ is$^6$
$$
\P (c_j ) = {{| \langle b|U(t_2 , t)|c_j \rangle ~\langle c_j |U(t, t_1 )
|a\rangle |^2} \over {\sum_k | \langle b|U(t_2 , t)|c_k \rangle ~\langle c_k
|U(t, t_1 ) |a\rangle |^2}}~~~~.
\eqno(17)$$
Still, the ABL formula applies to {\it ideal} intermediate measurements. Eq.
(17)
presupposes that the measurement of $C$ yields one of its eigenvalues $c_j$.
Real measurements, on the other hand, are subject to error. At the same time,
the disturbance they make is bounded.  These two aspects of real measurements
go together.  Suppose our measuring device interacts very weakly with the
systems in the ensemble.  We pay a price in precision.  On the other hand, the
measurements hardly disturb the ensemble, and therefore they characterize the
ensemble during the whole intermediate time.  Even non-commuting operators can
be measured at the same time if the measurements are imprecise. When such
measurements are made on pre- and post-selected ensembles, they yield
surprising results.  An operator yields {\it weak} values that need not be
eigenvalues, or even classically allowed.$^{1,7}$  The negative kinetic energy
of the previous section is an example of a weak value.

     Let us briefly review how weak values arise. The initial wave function of
the measuring device is $\Phi_{in} (Q)$.  After an impulsive measurement of an
operator $C$ and projection onto a final state, the final state of the
measuring device is
$$
\langle b |e^{-iPC/\hbar} |a\rangle \Phi _{in} (Q)
= \sum_i \langle b |c_i \rangle \langle c_i | a \rangle \Phi_{in} (Q-c_i )~~~~.
\eqno(18)$$
If $\Phi_{in} (Q)$ is sharply peaked, then the various terms $\Phi_{in} (Q-c_i
)$ will be practically orthogonal, and the probability of obtaining $c_i$ as an
outcome follows the ABL formula, Eq. (17).
But suppose $\Phi (Q)$ has a width $\epsilon$. Its Fourier transform has a
width in $P$ of $\hbar /\epsilon$. Small $\vert P \vert$ corresponds to a
measuring device that is coupled weakly to the measured system. If $\epsilon$
is large, then $\vert P \vert$ is small, and we can expand the exponential in
Eq. (18)
to first order in $P$ to obtain
$$\eqalign{
\langle b |e^{-iPC/\hbar} |a\rangle \Phi (Q)
&\approx  \langle b |1 {-iPC/\hbar} |a\rangle \Phi (Q)\cr
&\approx  \langle b |a\rangle e^{-iP C_w /\hbar} \Phi (Q)~~~~.\cr
}
\eqno(19)$$
Here
$$
C_w \equiv {{\langle a |C|b \rangle} \over {\langle a|b\rangle}}
\eqno(20)$$
is the {\it weak value} of the operator $C$ for the pre- and post-selected
ensemble defined by $\langle b|$ and $|a\rangle$.

     The definition of a weak value provides us with a new and intuitive
language for describing quantum processes.  In our example, the operators of
total energy $E$, kinetic energy $K$, and potential energy $V$ do not commute.
Therefore, the classical formula $E=K+V$ does not apply to the quantum
operators $E$, $K$, and $V$, but only to their expectation values; and the
expectation value of $K$ in any state is positive.  However, the formula
applies to weak values, as follows immediately from the definition, Eq. (20):
$$
 E _w = K_w +  V_w
{}~~~~,\eqno(21)$$
and the weak value of $K$ is {\it not} necessarily positive.  We can compute it
as $ K_w =  E_w -  V_w$.  We know $ E_w =E_0=-\alpha^2 \hbar^2 /2m$, since the
pre-selected state is an energy eigenstate, and $V_w$ vanishes since the
post-selected state is far from the potential well.  Then $K_w = -\alpha^2
\hbar^2 /2m$, the ``unphysical" obtained above in our example!  Weak values do
not appear in the conventional formulation of quantum mechanics, but they
appear in measurements.

     Eq. (19)
shows how weak values emerge from an imprecise measurement \hbox{($\de$
large)}.  But the weak value emerged from a {\it precise} measurement of
kinetic energy in our example. Instead of the condition on the initial state of
the measuring device ($\de$ large), we had a condition on the final state of
the particle ($x_0$ large and $\ep > \alpha \hbar^2 /m\de$). What do these
measurements have in common? Eq. (19)
assumes a weak measurement interaction which disturbs the measured system
within limits.  When $\vert P \vert$ is small, the measurement hardly intrudes
between the pre- and post-selected states of the system.  The pre- and
post-selected states define the measured system during the intervening time.
But when $\vert P \vert$ is not small, we can still control the effect of a
measurement.  In our example, we pre-select a state with negative total energy
and post-select a state where the potential vanishes.  It is not enough to
post-select particles outside the well.  The kinetic energy measurement
disturbs the particles, and they may not remain bound. We must somehow
post-select particles so far from the well that measurements of kinetic energy
could not have kicked them there. Then both negative total energy and vanishing
potential will characterize the particles throughout the measurement.

     To see what to post-select, let us write Eq. (12)
as an integral over $x$ instead of over $p$:
$$
\Phi_{fin} (Q) = \int_{-\infty}^{\infty} dx e^{-(x-x_0 )^2 /2\ep^2 }
e^{-{i\over \hbar} P p^2 /2m}  e^{-\alpha \vert x \vert } \Phi_{in} (Q)
{}~~~~,\eqno(22)$$
up to normalization.  If we could ignore the part of the integral near $x = 0$,
we could replace $p^2$ with $-\alpha^2$ in Eq. (22),
and the final state of the measuring device would be $\Phi_{fin} (Q) =
\Phi_{in} (Q+\alpha^2 \hbar^2 /2m)$.  Although we cannot ignore this part of
the integral, we can choose $\Psi_{fin} (x)$ to suppress it. $\Psi_{fin} (x)$
will suppress the integral near $x=0$ if Eq. (11) holds
and $\ep > \alpha \hbar^2 /m\de$.  We have already derived these conditions
(see Appendix).  Now we show intuitively, using time symmetry, how they keep
particles away from the well.

     Defining an ensemble via an initial state breaks time symmetry.  To
preserve time symmetry, we may select both an initial and a final state, thus
defining a pre- and post-selected ensemble. Both the ABL formula, Eq. (17),
and the definition of a weak value, Eq. (20),
manifest time symmetry.  We may think of quantum states propagating forwards
and backwards in time.$^7$  The initial state evolves forwards in time, and by
time symmetry the final state evolves backwards in time; {\it both} states
influence an intermediate measurement.  Indeed, the adjoint of Eq. (22)
represents reversed time evolution with $\Psi_{fin} (x)$ as the pre-selected
state and $\Psi_{in} (x)$ as the post-selected state.  If we reverse the time
evolution, the weak value remains the same, as well as the condition on
$\Psi_{fin} (x)$.  Applying the time evolution operator to $\Psi_{fin} (x)$,
$$
e^{ {i \over \hbar} P p^2 /2m } e^{- (x - x_0 )^2 /2 \ep^2 }
=\left( 1 -i \hbar P /m\ep^2 \right)^{-1/2}
e^{- (x - x_0 )^2 /2 ( \ep^2 - i\hbar P/m ) }
{}~~~~,\eqno(23)$$
we see that the effect of the measurement is to broaden $\Psi_{fin} (x)$. While
time evolution of $\Psi_{in} (x)$ can kick particles out of the well, time
evolution of $\Psi_{fin} (x)$ can bring particles {\it to} the well.  {\it
Either} forward {\it or} backward time evolution of $\Psi_{fin} (x)$ can bring
particles to the well, although forward time evolution is more familiar.

     Eq. (23)
is awkward because $P$ is an operator.  For a given value of $\vert P\vert$,
the semiclassical probability for the measurement to bring a particle to the
well is the absolute square of Eq. (23) for $x=0$.
Thus, for any state $\Phi_{in} (Q)$ with $\vert P \vert$ strictly bounded, such
as $({\sqrt \de }/{\sqrt \pi }Q ) \sin (Q/\de )$, a sufficient condition on
$x_0$ is
$$
x_0 > >2\alpha( \ep^2 + \hbar^2 P^2 /m^2 \ep^2 )
{}~~~~,\eqno(24)$$
for all $P$.  However, the gaussian state $\Phi_{in} (Q)$ of Eq. (4)
includes Fourier modes with arbitrary $\vert P\vert$. Large $\vert P \vert$ are
suppressed, but for no $x_0$ are they suppressed altogether. In the state
$\Phi_{in} (Q)$, the probability of a given $P$ is
$$
{\de \over { \hbar \sqrt{\pi}}} e^{-P^2 \de^2 / \hbar^2}
{}~~~~.\eqno(25)$$
Folding this probability with the absolute square of Eq. (23),
we obtain
$$
{{\de} \over {\hbar \sqrt{\pi}} }
\int_{-\infty}^{\infty} dP~ {{e^{-x_0^2 /(\ep^2+ \hbar^2 P^2 /m^2 \ep^2 )
}} \over {(1 + \hbar^2 P^2 /m^2 \ep^4 )^{1/2}} } e^{-P^2 \de^2 /\hbar^2 }
{}~~~~.\eqno(26)$$
as the probability for the measurement to bring particles to the well. For
large $x_0$ the integral is dominated by large $\vert P \vert$; we may replace
$\ep^2+ \hbar^2 P^2 /m^2 \ep^2$ by $\hbar^2 P^2 /m^2 \ep^2$ and neglect the
denominator to get an upper bound
$$
e^{-2x_0 m \ep \de /\hbar^2}
{}~~~~.\eqno(27)$$
If we {\it pre}-select $\Psi_{fin} (x)$, Eq. (27)
represents the fraction of the pre-selected ensemble that we would expect to
find at the well.  But the probability to {\it post}-select $\Psi_{in}$ is
suppressed by a factor $e^{-2\alpha x_0}$, for large $x_0$.  We want a pre- and
post-selected ensemble dominated by particles outside the well, and so we
require the latter probability to be much larger than the former: that is,
$$
\ep  > \alpha \hbar^2 /m \de
{}~~~~,\eqno(28)$$
with $x_0$ large.  These are the conditions imposed on $\Psi_{fin} (x)$ in
Section 3.  We need {\it both} conditions to restrict particles to the
classically forbidden region.  When these conditions hold, $V_w$ vanishes, and
a kinetic energy measurement yields $K_w$, even though the measurement is
precise.

     This is an important lesson:  the right pre- or post-selection allows us
to increase the precision of the intermediate measurement.  The price is that
we must wait for increasingly rare events.  As measurements of kinetic energy
become more precise ($\epsilon \rightarrow 0$), they disturb the particle more.
To get negative kinetic energies, we must post-select particles further from
the potential well ($x_0 \rightarrow \infty$). As the precision of the
measurement increases, negative kinetic energies become less and less frequent;
in the limit of ideal measurements, the probability vanishes, and so ideal
measurements of kinetic energy never yield negative values. It is easy to see
that if $\de$ approaches 0 while $x_0$ is held fixed, so does the chance to
measure negative kinetic energies. Taking the limit $\de \rightarrow 0$ in
Eq. (14)
turns $\Phi_{in} (Q-p^2 /2m)$ into a delta-function, and the final state of the
measuring device becomes (up to normalization)
$$
{\cos \left( \sqrt{2mQ} x_0 / \hbar \right)} \over
{\sqrt{2mQ}( \alpha^2 \hbar^2 + 2mQ)}
{}~~~~,\eqno(29)$$
for positive $Q$, and zero for negative $Q$.  The ABL formula predicts exactly
this distribution of kinetic energies.
\goodbreak
\bigskip
\noindent{\bf 5. \quad Conclusions}
\medskip
\nobreak
     We have seen that measurements of the kinetic energy of a particle in a
potential well can yield negative values consistently.  These measurements
involve selecting a final state of the particle far from the well. The negative
values represent the {\it weak} value of the kinetic energy operator.

     From the point of view of standard quantum theory, all that we have
produced is a game of errors of measurement. Ideal measurements of kinetic
energy can yield only positive values, since all eigenvalues of the kinetic
energy operator are positive. But in practice, measurements are not exact, and
even if their precision is very good, sometimes -- rarely -- they yield
negative values.  We have seen that if particles are subsequently found far
from the potential well, the measured kinetic energy of these particles comes
out negative.  Consistently, large measurement ``errors" did occur, producing a
distribution peaked at the ``unphysical" negative value $E_0$.  Mathematically,
this peak arises from an unusual interference. The measuring procedure pairs
each particle eigenstate of kinetic energy $K$ with a gaussian wave packet
$\exp [-(Q -K )^2 /2\de^2 ]$ of the pointer. But after projection onto the
post-selected particle state, these gaussians in $Q$ interfere, destructively
for positive $K$ and constructively for negative $K$.  The pointer is left in a
gaussian state centered on the negative value $E_0$, with a spread
characteristic of the measuring device.

     What special properties of non-ideal measurements led to this result?
First, these measurements involve only bounded disturbances of particle
position. Second, since their precision is limited, they can supply, ``by
error", the necessary negative values. These two properties are intimately
connected: any measurement of kinetic energy causing only bounded changes of
position  must occasionally yield negative values for the kinetic energy.  The
von Neumann formalism states that the change of $x$ due to the measurement is
$$\dot x={i\over {\hbar }}~g(t)~ [ x,~ P~ {p^{2}/ {2m}}]~~~~.
\eqno(30)$$
$P$ and $p$ are unchanged during the measurement, so the normalization
condition, Eq. (2), implies
$$x_{fin}-x_{in}=P~ {p/ {m}}~~~~.
\eqno(31)$$
{}From here it follows that the change of $x$ is bounded only if the pointer is
in an initial state with $P$ bounded, i.e. if the Fourier transform of
$\Phi_{in} (Q)$ has compact support. But then the support of $\Phi_{in}(Q)$ is
unbounded,$^8$ which immediately implies a nonzero probability for the pointer
to indicate negative values ($Q <0 $). Indeed, the ``game of errors" displays a
remarkable consistency, and this consistency allows negative kinetic energies
to enter physics in a natural way.

     The concept of a {\it weak} value of a quantum operator gives precise
meaning to the statement that the kinetic energy of a particle in a classically
forbidden region is negative:  namely, the weak value of the kinetic energy is
negative. Weak values are defined on pre- and post-selected ensembles.  The
interpretation of this concept raises subtle questions about time. Our example
involves pre-selection of particles in a bound state, measurement of their
kinetic energy, and post-selection of the particles far from the potential
well.  We associate negative values of kinetic energy with the particles.
However, instead of post-selecting particles far from the well, we could
measure the kinetic energy again with greater precision. We would then find
that almost every time the first measurement yielded a negative value, the
second measurement yields a positive value, and we would interpret the negative
value as an error of the measuring device.  The final measurement -- whether of
position or of kinetic energy -- is made {\it after} a kinetic energy
measurement has yielded negative values.  Nevertheless, the interpretation of
these negative values depends on the final measurement.  If we measure
position, we attribute them to the particle, while if we measure kinetic
energy, we attribute them to the device.  The effect seems to precede the
cause.

     The example of a particle in a potential well is a limiting case of
quantum tunnelling, when the barrier becomes very broad.  Negative kinetic
energies arise in finite barriers, too; but precise measurements of kinetic
energy require post-selected states deep in the classically forbidden region,
so negative kinetic energies may be hard to observe in narrow barriers.
Finally, we note a surprising extension to our result. By assuming an impulsive
measurement of kinetic energy, we could neglect the Hamiltonians of the system
and measuring device, and consider just their interaction. It follows that we
can observe particles with negative kinetic energy even if there is no binding
potential at all. What matters is only the shape of the pre-selected particle
wave function. Here, too, negative energies are consistent with other physical
processes (scattering).$^{9}$
\goodbreak
\bigskip
\centerline{\bf \quad Acknowledgement}
\medskip
\nobreak
S. P., D. R., and L. V. thank the Physics Department of the University of South
Carolina for hospitality during part of the writing of this paper.  D. R.
acknowledges support from the U.S.-Israel Binational Science Foundation through
the Hebrew University of Jerusalem, and from the Program in Alternative
Thinking at Tel-Aviv University. The research was supported by grant 425/91-1
of the the Basic Research Foundation (administered by the Israel Academy of
Sciences and Humanities) and by National Science Foundation grant PHY-8807812.
\goodbreak
\bigskip
\centerline{\bf \quad Appendix}
\medskip
\nobreak
We wish to obtain an upper bound for the magnitude of correction term,
Eq. (16),
in the final state of the measuring device. The absolute value of the
denominator is at least $\L ^2 - \alpha^2 \hbar^2$, so a bound is
$$
{{\hbar \alpha}\over \pi}
{{e^{(\alpha-\L /\hbar )x_0 -\alpha^2 \delta^2 /2}}
\over {\L ^2 - \alpha^2 \hbar^2} }
\int dp~~ e^{-(p^2 -\L ^2 ) \delta^2 /2 \hbar^2}
\vert \Phi_{in} \left( Q- (p-i\L )^2 /2m \right) \vert
{}~~~~,\eqno(A1)$$
and $\vert \Phi_{in} \left( Q- (p-i\L )^2 /2m \right) \vert$ is
$$
(\de^2 \pi )^{-1/4}
e^{-(Q+\L ^2 /2m)^2 /2\de} e^{ -{{p^4 } / {8m^2 \de^2}} +
p^2 ( {Q / {2m \de^2}} + {{3\L ^2} / {4 m^2 \de^2} }) }
{}~~~~.\eqno(A2)$$
Using$^{10}$
$$
\int_{-\infty}^{\infty} dp e^{-\mu p^4 \pm \vert a \vert
p^2} ={{\pi \vert a \vert}
 \over {4\mu }}  e^{a^2 /8\mu} \left[ I_{-{1 \over 4}}
\left( {{a^2} \over {8\mu}} \right) \pm
I_{1 \over 4} \left({{a^2} \over {8\mu}} \right) \right]
{}~~~~, \eqno(A3)$$
and
$$
I_{\nu} (x) = {{e^x }\over {(2\pi x)^{1/2}}} \left[ 1+ {\cal O} \left(
{1 \over x} \right) \right]
{}~~~~\eqno(A4)$$
for large $x$, we find that Eq. (A1)
leads to an exponential in
$$
(\alpha - \L /\hbar ) x_0
-\delta^2 \left( {{\L ^2} \over {\hbar^2}}  + {{\alpha^2}\over 2} \right)
+Q \left( {{\L ^2} \over {m\epsilon ^2}} - {{\delta^2 m} \over
{\hbar^2}} \right) + {{\L^4} \over {m^2 \epsilon^2}} + {{\delta^4 m^2
\epsilon^2} \over {2 \hbar^4 }}
{}~~~~.\eqno(A5)$$
The upper bound on the correction, Eq. (16),
will be exponentially suppressed if this sum of terms is sufficiently negative.
The parameter $\L$ is arbitrary, aside from the constraint $\L > \alpha
\hbar$. Since $\delta > \alpha \hbar^2 /m\de$ is a condition on $\Psi_{fin}
(x)$, we can eliminate the dependence on $Q$ by choosing $\L = \delta m \de
/\hbar $. Then for large enough $x_0$, the exponent is negative.  Setting $\ep
= n \alpha \hbar^2 /m\de$ for $n>1$, we obtain for the exponent
$$
-\alpha (n-1) x_0 +2({n^4 -n^2} )
{{\alpha^4 \hbar^4}\over {4 m^2 \de^2}}
{}~~~~,\eqno(A6)$$
so that the upper bound on the correction term is exponentially suppressed if
$$
\alpha x_0 >>  \left( {{\vert E_0 \vert }\over { \epsilon}} \right)^2
{}~~~~,\eqno(A7)$$
as in Eq. (11).
\goodbreak
\bigskip
\centerline{\bf \quad References}
\medskip
\nobreak
1.  Y. Aharonov, D. Albert and L. Vaidman, {\it Phys. Rev. Lett.} {\bf 60}
(1988) 1351.

2.  Y. Aharonov, J. Anandan, S. Popescu, and L. Vaidman, {\it Phys. Rev. Lett.}
{\bf 64} (1990) 2965.

3.  J. von Neumann,  {\it Mathematical Foundations of Quantum Theory}
(Princeton, New Jersey:  Princeton University Press, 1983).

4.  An example involving a smooth potential $V(x)
=-\alpha^2 \hbar^2 /m \cosh^2 (\alpha x)$ is presented in
Y. Aharonov, S. Popescu, D. Rohrlich, and L. Vaidman, in the Proceedings
of the International Symposium on the Foundations of Quantum Mechanics,
Tokyo, 1992, to appear.

5.  Y. Aharonov, P. G. Bergmann, and J. L. Lebowitz, {\it Phys. Rev.}
{\bf B134} (1964) 1410.

6.  If $C$ has degenerate eigenvalues, the projectors $\vert c_k \rangle
\langle c_k \vert$ appearing in Eq. (17)
must be replaced by projectors onto the degenerate eigenspaces.  See
Y. Aharonov and L. Vaidman, {J. Phys. A: Math. Gen.} {\bf 24} (1991) 2315.

7.  Y. Aharonov and L. Vaidman, {\it Phys. Rev.} {\bf A41} (1990) 11;
Y. Aharonov and D. Rohrlich, in {\it Quantum Coherence} (Proceedings
of the Conference on Fundamental Aspects of Quantum Theory, Columbia,
South Carolina, 1989), ed. J. S. Anandan (World-Scientific, 1990).

8.  If the Fourier transform of $\Phi _{in} (Q)$ has compact
support, then $\Phi_{in} (Q)$ is analytic. The two derivations of
our result, via contour integration and via Taylor expansion
of the exponential in Eq. (19),
both require $\Phi_{in} (Q)$ to be analytic.

9.  Y. Aharonov et al., Tel-Aviv University preprint TAUP 1847-90 (1991).

10.  I. S. Gradshteyn and I. M. Rhyzik, {\it Table of Integrals, Series, and
Products}, trans. and ed. A. Jeffrey (New York:  Academic Press, 1980).
\bye